\documentclass{amsart}

\usepackage{epsfig}
\usepackage{latexsym}
\usepackage{amsmath}
\usepackage{mathrsfs}
\newtheorem{theorem}{Theorem}[section]

\newtheorem{lemma}[theorem]{Lemma}

\newcommand{\kk}{l}

\newcommand{\T}{\mathcal T}

\newcommand{\PP}{{\mathbb P}}

\newcommand{\D}{{\mathscr D}}
   
\newcommand{\old}[1]{{}}

\title[Limits on building evolutionary trees]{A basic limitation on inferring phylogenies by pairwise sequence comparisons}
\author{Mike Steel}



\begin{document}

 \begin{abstract}
Distance-based approaches in phylogenetics such as Neighbor-Joining are a fast and popular approach for building trees. These methods take
pairs of sequences from them construct a value that, in expectation, is additive under a stochastic model of site substitution. Most models assume a distribution of rates across sites, often based on a gamma distribution. Provided  the (shape) parameter of this distribution is known, the method can correctly reconstruct the tree. However,  if the shape parameter is not known then we show that topologically different trees, with different shape parameters and associated positive branch lengths, can lead to exactly matching distributions on pairwise site patterns between all pairs of taxa.  Thus,  one could not distinguish between the two trees using pairs of sequences without some prior knowledge of the shape parameter.  More surprisingly, this can happen for {\em any} choice of distinct shape parameters on the two trees, and thus the result is not peculiar to a particular or contrived selection of the shape parameters. On a positive note, we point out known conditions where identifiability can be restored (namely, when the branch lengths are clocklike, or if methods such as maximum likelihood are used).
\end{abstract}
\maketitle

Allan Wilson Centre for Molecular Ecology and Evolution,

Biomathematics Research Centre,

University of Canterbury,

Christchurch,

New Zealand

Email:  m.steel@math.canterbury.ac.nz

\newpage
\bigskip
\section{Introduction}

Stochastic models that describe the evolution of aligned DNA sequence sites are fundamental to most modern approaches to phylogenetic tree reconstruction \cite{fel}. Making these models more realistic usually requires introducing additional parameters.    However, this raises the prospect that one might lose the ability to estimate a tree if one has to rely on the data to estimate all the parameters in the model. This could occur for various reasons -- for example, it may be that two different trees could produce exactly the same probability distribution on site patterns for two appropriately selected settings of the other parameters in the model. Such a scenario would be a problem for any method of tree reconstruction (including maximum likelihood and Bayesian methods) as it would mean that in some cases, one could not distinguish between two trees even with infinitely long sequences.  This loss of statistical `identifiability' has  been demonstrated for certain types of DNA substitution models, including rates-across-sites models \cite{szek} and, more recently, simple mixture models \cite{mat}.  On the positive side,  a number of identifiability results have also been established for suitably constrained models  (see, for example, \cite{all, all1, all2, all3, cha, ste1, ste}).

In this paper, we are interested in a phenomenon that is related to, but different from the loss of statistical identifiability, since it is method-dependent.  We will describe a situation where  pairwise sequence comparison methods  can fail to distinguish between trees, even though more sophisticated methods such as ML can.
Thus, the models are statistically identifiable, as far as the tree parameter is concerned,  but only if one uses the full matrix of aligned sequence information and not just pairwise sequence comparisons.  Specifically we consider tree reconstruction when sequences sites evolve under a model in which site rates have a gamma distribution, but where the (shape) parameter of the gamma distribution is not known. In this case, if one uses all the aligned sequence data, or at least $3$--way sequence comparisons then, for DNA sequences one can recover the shape parameter in a statistically consistent way, and thereby the underlying phylogenetic tree, by a recent result of Allman and Rhodes \cite{all}.  However if one just uses pairwise sequence comparisons we show that that two different trees can produce exactly the same pairwise sequence comparisons; moreover this can happen for any different choice of shape parameters for the two trees (by selecting the branch lengths on the two trees appropriately).

The intuition behind this limitation on pairwise sequence comparisons has been nicely summarized by Felsenstein (\cite{fel}, p. 175):  the rate at which a site is evolving affects all the taxa, but this constraint is not reflected by a method that is based on pairwise comparisons, and so, for example,  ``once one is looking at changes within rodents it will forget where changes were seen among primates."

Before describing our results we mention some earlier papers that described related by different phenomena.  Baake \cite{baa}  considered a model in which half the sites are invariable and the remaining sites evolve under a general Markov model. Although this model (and the tree) is generically identifiable using all the sequence information (as recently shown in \cite{all2}) Baake showed that two trees can produce identical pairwise sequence comparisons.    The non-indentifyability of divergence times on a fixed tree under various rates-across-sites models has also been recently investigated by Evans and Warnow \cite{eva}. Finally we note that our result that distance-based methods can be misleading for tree inference complement some earlier work \cite{ban}, \cite{hus} which highlighted a different result in which distances can perfectly `fit' one phylogenetic tree when the full sequence data support a different tree.

\section{Definitions and observations}

In sequence-based approaches to phylogenetics, the data usually consists of a collection of $n$ sequences  $s^1, s^2, \ldots, s^n$, each of length $N$, and where each sequence site takes values in some state space. We will suppose that there are $r$ states, and denote them by greek letters $\mu, \nu$ throughout - for example, for aligned DNA sequence data $r=4$
and the state space is the four DNA bases (A,C,G,T).  Given the aligned sequences, biologists seek to infer a phylogenetic tree $\T$, whose leaves are labeled by $\{1,\ldots, n\}$ and which describes the evolution of the sequences from some unknown common ancestral sequence (leaf $i$ corresponds to the extant taxon from which sequence $s^i$ has been obtained).
For further background on phylogenetics, the reader may consult \cite{fel, sem}.

Given two sequences $s^i= (s_1, s_2, \ldots, s_N)$  and $s^j= (s_1', s_2', \ldots, s_N')$ let  $\hat{J}_{ij}$ be the $r \times r$ matrix whose $\mu\nu$--entry is the proportion of sites where sequence $s^i$ is in state $\mu$ and sequence $s^j$ is in state $\nu$.  The proportion of sites where sequence $s^i$ and $s^j$ differ, $\delta_{ij}$ (the normalized sequence dissimilarity) is therefore the sum of the off-diagonal entries of $\hat{J}_{ij}$; more formally,
$\delta_{ij} = \frac{1}{N}\sum_{k=1}^N\{k: s^i_k \neq s^j_k\}=1-tr(\hat{J}_{ij}),$
where $tr$ refers to matrix trace (the sum of the diagonal entries).

Given a collection of sequences $s^1, s^2, \ldots, s^n$, each of length $N$, one can easily derive the collection of pairwise $\hat{J}$--matrices  $\hat{J}_{ij}: i,j \in \{1, \ldots, n\}$.
This  reduction process, from aligned sequences to pairwise comparisons,  is highly redundant (for typical values of $n$) since it reduces the frequencies of $r^n$ site patterns to $\binom{n}{2}$ comparisons of $r^2$ sites pattern frequencies. The further reduction to the $\delta$ values involves even more redundancy \cite{ste88}.  Despite this, it is well known that these reduced matrices (and sometimes just the $\delta$ values) provide a statistically consistent way to estimate the underlying tree,  under simple models of DNA site substitution. This follows by combining two well-known facts.

\noindent{\bf Fact One: }  Under the assumption that the aligned sequence sites evolve i.i.d., the law of large numbers tells us that the $\hat{J}_{ij}$ matrices  (and thereby the $\delta_{ij}$ values)  converge in probability to their expected values as the sequence length $N$ becomes large.

To explain this further we introduce two key definitions:  For $i, j \in X$, let $J_{ij}$ be the expected value of $\hat{J}_{ij}$ -- thus, $J_{ij}$ is an $r \times r$ matrix whose $\mu\nu$--entry  is $$J_{ij}^{\mu\nu}:= \PP(s_k^i = \mu, s_k^j = \nu)$$
for each pair of states $\mu, \nu$, and any given $k$; and let
 $$d_{ij}: = \PP(s_i^k \neq s_j^k),$$ for any given $k$.
In words, $J_{ij}$ is the matrix whose entries describe the joint probability that at any given site the sequences $s^i$ and $s^j$ are in specified states, while $d_{ij}$ is simply the probability that these states are different at a given site. By definition, $d_{ij} = 1- tr(J_{ij}).$

With this notation, Fact One can be restated as the condition that, for all $i,j \in X$:
$$\hat{J}_{ij}  \overset{p}{\rightarrow} J_{ij} \mbox{   and   }  \delta_{ij}  \overset{p}{\rightarrow} d_{ij},$$
where $ \overset{p}{\rightarrow}$ denotes convergence in probability as $N \rightarrow \infty$.

The second result required to show that the $\hat{J}_{ij}$ values estimate the tree consistenty is that for many models the $J_{ij}$ values can be transformed to obtain a function on pairs of leaves that is additive.  Recall that  a function $l_{ij}$ on pairs of leaves of a tree is said to be {\em additive} on a tree $\T$  if one can assign a positive real number $l_e$ to each edge $e$  of $\T$ so that $l_{ij}$ is the sum of the numbers assigned to the edges on the path connecting the two leaves on the tree.  That is:
\begin{equation}
\label{lin}
\kk_{ij} = \sum_{e \in p(\T; i,j)} \kk_e,
\end{equation}
 where $p(\T;i, j)$ denotes the edges on the path in $\T$ connecting $i$ and $j$.
This additivity condition implies that  the tree $\T$ can be uniquely recovered from the $ l_{ij}$ values  (see e.g. \cite{sem}).  With this in mind we have:

\noindent{\bf Fact Two: }  Under various models of sequence site evolution, a distance function $l$ on $X$ that is additive on the underlying tree can be computed from the $J$  matrices (and sometimes just the $d$ values).

The two main models for which Fact Two is known to apply are (i) the general Markov process, for which the transformation $J_{ij} \mapsto -\log(\det(J_{ij}))$ is additive, and (ii)
the general time-reversible (GTR) model with any  known distribution of rates across sites. In this latter case  -- which is the one of interest in this paper -- one can transform the $J$ matrices to obtain an distance function $l$ on $X$ that corresponds to the expected number of substitution (`evolutionary distance') between $i$ and $j$ -- and which is therefore additive.
For a GTR model, with a distribution $\D$ of rates across sites this transformation \cite{wad} is:
$$l_{ij} = -tr(\Pi M_{\D}^{-1}(\Pi^{-1}J_{ij})),$$
where $M_{\D}$ is the moment generating function of the distribution of rates across sites, and where $\Pi = {\rm diag}(\pi)$ is the diagonal matrix whose leading diagonal is the vector
 $\pi =[\pi_\mu]$ of the frequencies of the $r$ states.
For the GTR model (or any submodel) the matrix $J_{ij}$ is symmetric \cite{wad} and $J_{ii} = \Pi$ for each $i$.

Combining Fact One and Fact Two gives:
$$-tr(\Pi M_{\D}^{-1}(\Pi^{-1}\hat{J}_{ij})) \overset{p}{\rightarrow} l_{ij}.$$
and so the $\hat{J}_{ij}$ values allow us to reconstruct the underlying tree from sufficiently long sequences.
Indeed even if we don't know the stationary frequencies of the states (the matrix $\Pi$) we can still recover the tree, since $\Pi$ is determined by (the row sums of) $J_{ij}$, and so if we
let $\hat{\Pi}_{ij}$ denote the corresponding empirical state frequencies (determined by the corresponding row sums of $\hat{J}_{ij}$) then we have:
$$-tr(\hat{\Pi} M_{\D}^{-1}(\hat{\Pi}^{-1}\hat{J}_{ij})) \overset{p}{\rightarrow} l_{ij}.$$

Thus, if for each pair $i,j$ we derive an estimate $\hat{l}_{ij}$ of evolutionary distance ($l_{ij}$) by either maximum likelihood estimation or by the `corrected distance' formula:
\begin{equation}
\label{lij}
\hat{l}_{ij} = -tr(\hat{\Pi} M_{\D}^{-1}(\hat{\Pi}^{-1}\hat{J}_{ij}))
\end{equation}
then these estimated values will converge to the true $l_{ij}$ values as the sequence length $N$ grows, allowing for statistically consistent reconstruction of the tree by using fast distance-based tree reconstruction methods.

For some GTR models it is also possible to transform just the $\delta_{ij}$ to obtain $l_{ij}$ -- for example,  under
the simple symmetric 4-state model (the Jukes-Cantor model) the transformation is:
\begin{equation}
\label{cor1}
l_{ij} = -\frac{3}{4}M^{-1}_{\D}(1-\frac{4}{3}d_{ij}).
\end{equation}
For models in which $l_{ij}$ can be expressed as a function of $d_{ij}$ one can use $\delta$ in place of $d$ to estimate $l_{ij}$ (for certain models, such as the Jukes-Cantor model, this leads to the same $l_{ij}$ estimates as a pairwise maximum likelihood estimate, but for more complex models this need not be the case).

The snag  in this otherwise appealing story is that it assumes that we  know the distribution  $\D$ of rates across sites -- what happens if $\D$ is unknown or has parameters that require estimation?
If no constraints are placed upon $\D$ then identifiability of the tree can be completely lost \cite{szek}. It is therefore fortunate that in molecular systematics $\D$ is typically described by a simple parametric distribution. In particular,  the gamma distribution has a long and popular history in models that describe the variation  of substitution rates across DNA sequence sites \cite{yan}.  Today, a common default option is the  `GTR+$\Gamma$+I' model  in which each site is either invariant (with some probability), or it evolves according to a general time reversible Markov process that proceeds at a rate selected randomly from a gamma distribution.   In this paper we will ignore the invariable sites, since our main result (Theorem \ref {mainthm}) will automatically imply a corresponding result when invariable sites are present.   Moreover, we may (without loss of generality) assume that the gamma distribution is normalised so that its mean is equal to $1$ and so there remains just one parameter - the `shape' parameter, $k$.

We will show that {\em any} two different shape parameters can provide exactly the same $J$ matrices on a pair of topologically distinct trees (with appropriately assigned branch lengths). Consequently, using just pairwise comparisons (the $\hat{J}$ matrices) to infer phylogeny from the resulting data, without prior knowledge of the shape parameter is potentially problematic -- either of the two trees could describe the data much better than the other if one were to select the shape parameter appropriate for that tree.  Thus, a  biologist exploring data by seeing the effect of varying $k$ might note that for one value of $k$ his/her data fit a tree perfectly. The result described here shows that it could be dangerous to stop at this point and report the tree, as there may well be another value of $k$ for which the pairwise sequence data (or  distance data) fit a different tree perfectly.   Using all the data (i.e. not reducing to pairwise comparisons) will overcome this problem for a gamma distribution as established recently by Allman and Rhodes \cite{all} (who also pointed out errors in an earlier approach from \cite{rog}).

\section{Results}

In this paper we consider a particular type of reversible stationary markov process, called the {\em equal input model}.   In this model, the rate of substitution does not depend on the current state, and when a substitution event occurs, the new state is selected according to the stationary distribution of states, which we encode by the vector $\pi$.  Thus the rate matrix $R$ is defined by the condition $R_{\mu\nu} = \pi_{\nu}$ for all $\nu \neq \mu$.   In the case of $r=4$ states, this model has been called the `Tajima-Nei equal input model' or the `Felsenstein 1981 model'; when, in addition, $\pi$ is uniform, it is the known as the `Jukes-Cantor' model.  For more mathematical background on the equal input model, see, for example, \cite{sem}. Although the equal-input model is a special case of the GTR model, we have chosen it because it is simple enough to allow tractable exact calculations, yet without being overly simplistic (for example, it allows arbitrary stationary frequencies for the states).

 Under the equal input model, and with constant-rate site evolution we have:
\begin{equation}
J_{ij}^{\mu\nu}=
\begin{cases}
\pi_{\mu}\pi_{\nu}(1-\exp(-l_{ij}/\gamma)),  & \text { if }\mu \neq \nu ;\\
\pi_{\mu}(\pi_{\mu}+ (1-\pi_{\mu})\exp(-l_{ij}/\gamma))), & \text { if } \mu = \nu,
\end{cases}
\end{equation}
where $l_{ij}$ is the expected number of substitutions on the path connecting $i$ and $j$ in $\T$ (an additive distance) and
$\gamma = 1-\sum_{\mu}\pi_\mu^2$ (this number is the expected normalised sequence dissimilarity for saturated sequences -- for example, in the Jukes-Cantor model, it takes the value $1-4\cdot (\frac{1}{4})^2 = \frac{3}{4}$).
More briefly we can write:
\begin{equation}
\label{JJ}
J_{ij}^{\mu\nu}= a_{\mu\nu}+ b_{\mu\nu}\exp(-l_{ij}/\gamma),
\end{equation}
where $a_{\mu\nu}, b_{\mu\nu}$ are constants that depend on the pair $\mu, \nu$ and the vector $\pi$.

If we now  impose an associated distribution $\D$ of rates across sites on this equal-input model, in which case each site evolves according to the same equal input model, but with a rate selected randomly according to $\D$.  In this case (\ref{JJ}) becomes:
\begin{equation}
\label{dkap}
J_{ij}^{\mu\nu} =a_{\mu\nu} + b_{\mu\nu}M_\D(-l_{ij}/\gamma),
\end{equation}
where $M_\D(x)$ is the moment generating function for $\D$, and
When $\D$ is a gamma distribution of rates across sites with shape parameter $k$ and mean $1$ we have:
$$M_\D(x) = (1-\frac{x}{k})^{-k},$$
and so Eqn. (\ref{dkap}) becomes:
\begin{equation}
\label{dkap2}
J_{ij}^{\mu\nu} =  a_{\mu\nu} + b_{\mu\nu}(1 + \frac{\kk_{ij}}{k\gamma})^{-k}
\end{equation}

Now, suppose we have two topologically distinct binary phylogenetic $X$--trees $\T$ and $\T'$, where $\T$ has branch length $l$ and gamma distribution of rates across sites (with mean $1$) with shape parameter $k$, while $\T'$ has branch lengths $l'$ and gamma distribution of rates across sites (with mean $1$) with shape parameter  $k'$, where $k' \neq k$.
We can now state the main result of this paper.

\begin{theorem}
\label{mainthm}
Consider a fixed equal imput model on $r \geq 2$ states. Then
for any $k, k'>0$ with $k \neq k'$ and for any binary phylogenetic $X$--tree $\T$ with four or more leaves there exist a topologically distinct binary phylogenetic $X$--tree $\T'$, and strictly positive branch lengths $\kk$ for $\T$ and $\kk'$ for $\T'$ respectively,  so that the  matrices of joint pairwise distributions $J_{ij}$ and $J'_{ij}$ agree for all $i, j \in X.$
\end{theorem}

{\bf Remarks:} The significance of this result for phylogenetic reconstruction is that it shows that even if one uses pairwise sequence comparisons, the choice of the correct shape parameter for the gamma distribution is essential -- if we selected shape parameter $k$, the corrected distances (obtained by ML estimation or by (\ref{lij})) would fit $\T$ perfectly as the sequence lengths become large; while if we selected shape parameter $k'$, the corrected distances would fit $\T'$ perfectly for sufficiently long sequences.  Notice that the pair $(\T, k)$  and $(\T', k')$ fit the data produced by either tree (with its associated shape parameter) equally well (i.e. perfectly in the limit as the sequence lengths become large).  Moreover, our result
assumes that the base frequency vector ($\pi$) is known and the same for both trees. Notice also that Theorem \ref{mainthm} automatically implies that any distance correction method that transforms the sequences dissimilarities (the $\delta$ values) will  be unable to distinguish between $\T$ and $\T'$ if the shape parameter is unknown.

{\em Proof of Theorem~\ref{mainthm}:}

\noindent For a given assignment of branch lengths $l$ and gamma shape parameter $k$  for  $\T$ let $J_{ij}$ denote the induced pairwise distribution matrix, defined by Eqn.~(\ref{dkap2}), for each $i,j$.
Similarly, for a  given assignment of branch lengths $l'$ and gamma shape parameter $k'$  for  $\T'$ let $J'_{ij}$ denote the induced pairwise distribution matrix for each $i,j$.
By symmetry, we may assume (without loss of generality) that $k>k'$.
Let
\begin{equation}
\label{taudef}
\tau_{ij} := 1+\frac{ \kk_{ij}}{k\gamma},
\end{equation}
\begin{equation}
\label{taudef2}
\tau'_{ij} := 1+\frac{\kk'_{ij}}{k'\gamma},
\end{equation}
and let $$\rho =\frac{k}{k'}>1.$$
 From Eqn.~(\ref{dkap2}) and the notation of (\ref{taudef}) and (\ref{taudef2}), we have the following fundamental identity:
\begin{equation}
\label{thishelps}
J_{ij} = J'_{ij} \mbox{ if and only if  } \tau'_{ij}= (\tau_{ij})^\rho
\end{equation}
We will first prove Theorem~\ref{mainthm} in the case where $|X|=4$, and then extend the proof to the general case.

{\bf The case $|X|=4$:}  Consider the tree $\T$ with branch lengths given in Fig. 1(a), and the tree $\T'$ with branch lengths given in Fig. 1(b).  By (\ref{lin}) we have, for example,
$\kk_{12}= \kk_1 + \kk_2$, and $\kk_{13} = \kk_1+\kk_3+\kk_5$.
Let $\kk'_{ij}$ be the corresponding $\kk'$ values induced by $\T'$.
\begin{figure}[h]
\begin{center}
\resizebox{10.5cm}{!}{
\includegraphics {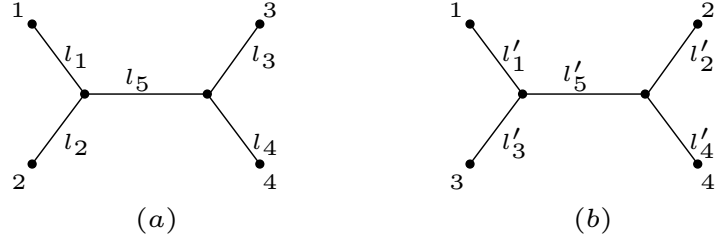}
}
\caption{(a) Tree $\T$ with branch lengths $l$; (b) Tree $\T'$ with branch lengths $l'$.}
\end{center}
\end{figure}

Notice that if we set
\begin{equation}
\label{xidef}
x_i := \frac{1}{2} + \frac{\kk_i}{k\gamma} \mbox{  for } i =1,\ldots, 4
\end{equation}
 and set
\begin{equation}
\label{epsdef}
\epsilon := \frac{\kk_5}{k\gamma}.
\end{equation}
Then for each distinct pair $i,j$ we have:
\begin{equation}
\label{cases}
\tau_{ij}=\begin{cases}
x_i+x_j, & \text { if } \{i,j\} =\{1,2\} \text{ or } \{3,4\} ;\\
x_i+x_j+\epsilon, & \text { otherwise.}
\end{cases}
\end{equation}
thus $\tau_{ij}$ is additive on $\T$ (similarly, $\tau_{ij}'$ defined by (\ref{taudef2}) is additive on $\T'$).

We will describe an assignment of positive branch lengths for $\T$, and then an assignment of branch lengths for $\T'$. Firstly, however we state a
convexity lemma; for completeness a proof is provided in the Appendix.
\begin{lemma}
\label{helpful}
Suppose $f$ is twice-differentiable, and that $f''$ is strictly positive on the positive reals.
If $u' \leq u \leq v \leq v'$ and $u+v = u'+v'$ then:
$ f(u')+f(v') > f(u)+f(v) .$
\end{lemma}
We will apply Lemma~\ref{helpful} twice during the proof, using the function $f(x) = x^\rho$ which satisfies the hypotheses of this lemma, since  $f''(x) = \rho(\rho-1)x^{\rho -2}$ and $\rho>1$.

Returning to the assignment of branch lengths for $\T$, let
$L>\frac{1}{2}$ and $t\in [0,1]$, and select $l_1, \ldots, l_5$ so that $(x_1, x_2, x_3, x_4, \epsilon)$  defined by (\ref{xidef}) and (\ref{epsdef}) satisfy the following system of inequalities:
\begin{equation}
\label{ineq1}
\max\{x_1,x_2, x_3,x_4\} = L
\end{equation}
\begin{equation}
\label{ineq2}
x_3 \geq x_4+t;
\end{equation}
\begin{equation}
\label{ineq3}
x_2<x_4;
\end{equation}
\begin{equation}
\label{ineq4}
 x_1+x_3\leq x_2+x_4;
 \end{equation}
 \begin{equation}
\label{ineq5}
 x_1+x_4 \leq x_2+x_3;
 \end{equation}
 \begin{equation}
\label{ineq6}
|x_i - x_j| \leq 1 \mbox{ for all $i, j$};
 \end{equation}
 and
  \begin{equation}
\label{ineq7}
(x_1+x_4+\epsilon)^\rho+(x_2+x_3+\epsilon)^\rho = (x_1+x_2)^\rho + (x_3+x_4)^\rho.
 \end{equation}

We pause to observe that this system (for the five $l_i$ values) is feasible for arbitrarily large values of $L$. For example, we can take
$x_1= L, x_2 = L+\frac{1}{3}, x_3=L+1, x_4= L+\frac{2}{3}$ and $t=\frac{1}{3}$, to satisfy (\ref{ineq1})--(\ref{ineq6}),
and then for $i =1, \ldots, 4$ let $l_i = k\gamma(x_i - \frac{1}{2})$, which is strictly positive since $x_i \geq L > \frac{1}{2}$; then for $l_5$ there exists a positive value of $\epsilon$ satisfying
(\ref{ineq7}). To see this last claim regarding $\epsilon$,  let  $$u = x_1+x_4; \mbox{  } v= x_2+x_3,$$ and
$$u' = x_1+x_2; \mbox{  }  v' = x_3+x_4.$$
Notice that the inequalities (\ref{ineq3}) and (\ref{ineq5})
imply that $u'<u\leq v<v'$ and, since $u+v=u'+v'$, Lemma~\ref{helpful} applied to $f(x) = x^\rho$ gives  $f(u)+f(v)<f(u')+f(v')$.  Since $f$ is strictly increasing, this implies that there is a finite and strictly positive value of $\epsilon>0$ (and thereby of $l_5$ by (\ref{epsdef})) for which
$f(u+\epsilon) + f(v+\epsilon) = f(u')+f(v')$, as claimed.

Next we show that the branch lengths we have assigned for $\T$ allows us to assign positive branch lengths to $\T'$ so $J_{ij}=J'_{ij}$ holds for all $i,j$.
Define $\lambda_{ij}: = f(\tau_{ij})$ where $f(x) = x^\rho$. We will show that there exists an assignment of positive branch lengths $l'$ to $\T'$ for which the associated vector $\tau'$ defined by  (\ref{taudef2}) satisfies:
\begin{equation}
\label{lama}
\lambda_{ij} = \tau_{ij}'.
\end{equation}
In view of (\ref{thishelps}) this will establish the theorem in the case $|X|=4$.  Let
\begin{equation}
\label{Seqs}
 S'_{12|34} := \lambda_{12}+\lambda_{34}, S'_{13|24}:=\lambda_{13}+\lambda_{24}, \mbox{  and  } S'_{14|23} := \lambda_{14}+\lambda_{23}.
\end{equation}
If we let  $$u = x_1+x_3+\epsilon; \mbox{  }   v= x_2+x_4+\epsilon,$$ and
$$u' = x_1+x_4+\epsilon; \mbox{  }  v' = x_2+x_3 +\epsilon,$$
then
(\ref{ineq2}) and (\ref{ineq4}) imply that  $u'<u\leq v<v'$ and, since $u+v=u'+v'$,  Lemma~\ref{helpful} gives
$f(u)+f(v)<f(u')+f(v')$.  In view of (\ref{Seqs}) and (\ref{cases}) this implies that:
\begin{equation}
\label{helps2}
S'_{13|24} < S'_{14|23}.
\end{equation}
Moreover, Eqn.~(\ref{ineq7}) implies that
\begin{equation}
\label{helps3}
S'_{12|34} = S'_{14|23}.
\end{equation}
Equations (\ref{helps2}) and (\ref{helps3}) imply that $\lambda_{ij}$ can be realized as a sum of real-valued branch lengths on $\T'$ by assigning  positive interior branch length (call it $\epsilon'$), and real-valued (possibly negative) pendant branch lengths (by \cite{hak}).  We will first show that these four pendant branch lengths are not only positive, but also strictly greater than $\frac{1}{2}$ provided $L$ is chosen sufficiently large.
For $i \in \{1,2,3,4\}$ if we let $\lambda_i$ denote the branch length of the edge incident with leaf $i$, then
$\lambda_i = \frac{1}{2}(\lambda_{ij} + \lambda_{ik}-\lambda_{jk})$ for any choice $j,k$ for which
$|\{i,j,k\}|=3$.
Now, from (\ref{ineq1}) and (\ref{ineq6}), we have
$$\lambda_{ij} + \lambda_{ik}-\lambda_{jk} \geq f(2L)+f(2L) - f(2L+2+\epsilon)$$ and so we
can select a value of $L$ that is sufficiently large to ensure that $\lambda_i >\frac{1}{2}$ for $i=1\ldots, 4$.
We can now assign the positive branch lengths to $\T'$ as follows.
Let $l'_5 = k'\gamma \epsilon'$ and for $i \in \{1,\ldots, 4\}$ let
$$l'_i = k'\gamma(\lambda_i -\frac{1}{2})>0.$$
With these branch lengths we have (from (\ref{taudef2}), (\ref{lama})),
$$\tau'_{ij} = \lambda_{ij}= (\tau_{ij})^\rho,$$ for all $i,j$.  By Eqn.~(\ref{thishelps}), this establishes the theorem in the case where $|X|=4$.

{\bf The case $|X|>4$:}  To extend the proof to larger trees we require a further lemma, which is based on the following definition.
Given a rooted phylogenetic tree, $t$ with root vertex $\rho$ (which we assume is a vertex of degree at least two) and associated branch lengths $l$, we say that the branch-lengths on $t$ are {\em clock-like} if the sum of the branch lengths from $\rho$ to any leaf takes the same value for each leaf, which we will denote by $h(t, l)$ (the `height' of $\rho$).
We will use the following lemma, for which a proof is provided in the Appendix.
\begin{lemma}
\label{ultralem}
Let $t$ be a rooted phylogenetic tree with at least two leaves.
Suppose that the branch-lengths for $t$ are clock-like, and that we have a gamma distribution of rates across sites (with mean $1$) and with shape parameter $k$.  For any other shape parameter $k'$  there exists a unique associated vector of branch lengths $l'$  for $t$  that are clock-like and such that the induced $J'$ matrices satisfy the condition:
 \begin{equation}
 \label{ultra1}
 J'_{ij}  = J_{ij} \mbox{   for all leaves $i, j$  of }  t.
  \end{equation}
 Moreover, for this vector $l'$, we have:
   \begin{equation}
 \label{ultra2}
  h(t, l') =\frac{1}{2}k'\gamma(-1+(1+\frac{2h(t,l)}{k\gamma})^\rho).
 \end{equation}
\end{lemma}
 Returning to the proof of the theorem, let $\T$ be any binary phylogenetic tree with more than four leaves, and select any interior edge $e$ of $\T$. Consider the four rooted subtrees $t_1, t_2, t_3, t_4$ of $\T$ that result from deleting this edge and its two endpoints, as shown in Fig. 2(a).
\begin{figure}[ht]
\begin{center}
\resizebox{12cm}{!}{
\includegraphics {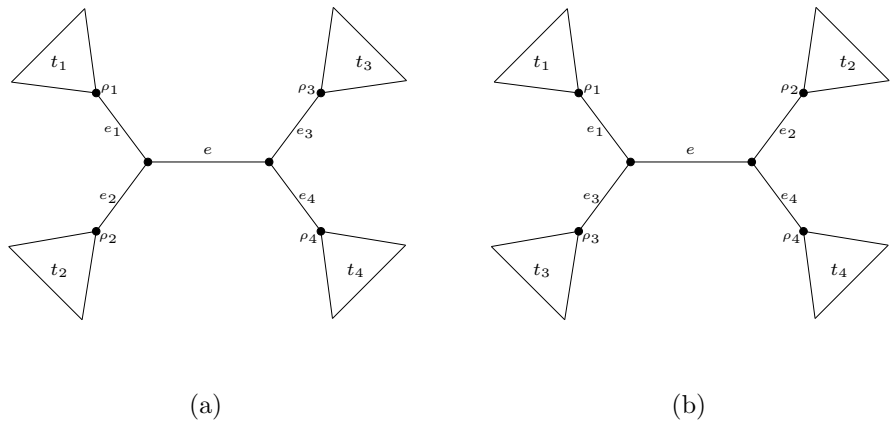}
}
\caption{Representation of $\T$ in (a), and of $\T'$ in (b), when $|X|>4$.}
\end{center}
\end{figure}
Let $\T'$ be the tree obtained from $\T$ by interchanging the subtrees $t_2$ and $t_3$, as shown in Fig. 2(b).  Let  $l$ and $l'$ be strictly positive branch lengths for the two quartet trees of Fig. 1 for which we have $J_{ij} = J'_{ij}$ for all $i,j \in \{1,2,3,4\}$ (by the case of the theorem established already for $|X| = 4$).

We now assign branch lengths to $\T$ and $\T'$.
For tree $\T$ assign length $l_5$ to edge $e$ indicated in Fig. 2(a), and for the tree $\T'$  assign length $l_5'$ to edge $e$ of $\T'$ indicated in Fig. 2(b).
  If $t_i$ consists of just a single leaf, we assign length $l_i$ in $\T$ and $l_i'$ in $\T'$.  Thus it remains to specify how we assign branch lengths to
  the subtrees $t_i$ when these trees contain more than one leaf, and to the edge $e_i$ that connects $t_i$ to  $e$.
If we regard $t_i$ as a rooted binary phylogenetic tree (for which the root $\rho_i$  is the vertex adjacent to an endpoint of $e$, as shown in Fig. 2), we  assign branch lengths to $t_i$ that are clock-like and for which
 $h(t_i, \rho) = \xi_i$, where $\xi_i$ is any strictly positive number that is less than $l_i$ and satisfying the condition:
 \begin{equation}
  \label{ultra3}
\frac{1}{2}k'\gamma(-1+(1+\frac{2\xi_i}{k\gamma})^\rho) < l_i'.
 \end{equation}
Then assign edge $e_i$ length $l_i - \xi_i >0$. Note that we can select $\xi_i$ to satisfy (\ref{ultra3}) since the left-hand side of (\ref{ultra3}) converges to zero as $\xi_i \rightarrow 0$.
 For tree $\T'$ assign $t_i'$ branch lengths that are clock-like and satisfy  (\ref{ultra1}) of Lemma \ref{ultralem} (for $t = t_i, t' = t_i'$), and
assign edge $e_i$ length $l_i' - h(t_i', \rho)$ which is strictly positive by (\ref{ultra3}).
 We claim that $J_{ij} = J'_{ij}$ for all $i,j$.  We have just shown that this holds whenever $\{i,j\}$ are leaves in the same subtree ($t_1, t_2, t_3$ or $t_4)$, thus it remains to check the claim when $i$ and $j$ lie in different subtrees, say $t_r, t_s$.  In this case the condition that $(\T, l)$ and $(\T', l')$ satisfy the theorem in the case $|X|=4$ and the fact that the distance
between $i$ and $j$ in $\T$ is $l_{rs}$ and in $\T'$ is $l'_{rs}$ (according to the way the branch lengths have been assigned) establishes case (ii).  This completes the proof.
\hfill$\Box$

\section{Concluding comments}

Our result shows that  rate variation across sites can indeed provide an ``inherent limitation that is worrisome'' \cite{fel} for methods that rely solely on pairwise sequence comparisons.
Despite the limitation of distance-based phylogenetic reconstruction imposed by Theorem \ref{mainthm}, there is one situation where distances suffice to recover a tree under a gamma rate distribution across sites, even when the shape parameter is unknown. This is when the underlying branch lengths on the tree obey are clock-like (i.e. obey a `molecular clock').  This follows from the monotone relationship between $d$ and $\kk$ described in (\ref{thishelps}), which implies that the $d$ values (corrected or not) will be ultrametric and additive on the underlying tree.

Also, our result does not imply that tree reconstruction is hopeless without prior or independent knowledge of the shape parameter, since Allman and Rhodes \cite{all} have established that identifiability holds for this model (generically for all $r \geq 2$, and exactly when $r=4$ which is the case that applies for DNA sequence data) and so methods such as maximum likelihood will be statistically consistent.  Moreover, their result shows that just $3$--way sequence comparisons are sufficient to identify the shape parameter. This suggests that it may be possible to develop statistically consistent but fast modifications of distance-based tree reconstruction methods (such as neighbor joining) that some allow triple-wise calculations.

Finally,  it would also be interesting to check whether Theorem \ref{mainthm} remains true if one replaces the equal input model by the GTR model with any fixed (and given) rate matrix $R$.   This seems quite likely, though the calculations appear to be more involved when the rate matrix has many different eigenvalues.  The question of whether $\T'$ can have an arbitrary topology different to $\T$ in Theorem \ref{mainthm} (i.e. not just a nearest-neighbor interchange of $\T$) could also be of interest.

\section{Acknowledgements} I thank Joe Felsenstein for several helpful comments, and whose talk at the Sante Fe Institute (April 2008) on a related problem motivated the present study.
This work is supported by the {\em Allan Wilson Centre for Molecular Ecology and Evolution}.


\section{Appendix}
{\bf Proof of Lemmas~\ref{helpful} and \ref{ultralem}}.

{\em Proof of Lemma~\ref{helpful}.}
By a Maclaurin series expansion, we have:
$$f(u') = f(u-t) = f(u) -tf'(u) +\frac{1}{2}t^2f''(\theta),$$ where $\theta \in [u', u]$ and:
$$f(v') = f(v+t) = f(v)+tf'(v) + \frac{1}{2}t^2f''(\theta'),$$ where $\theta' \in [v, v']$.
Thus: $$f(u')+f(v') = f(u)+f(v) + t (f'(v)-f'(u)) + \frac{1}{2}t^2(f''(\theta) + f''(\theta')).$$
Now $f'(v)-f'(u)>0$ since $f'$ is increasing (by the positivity of $f''$),  and so, since $t>0$:
$$ f(u')+f(v') >  f(u)+f(v)  + \frac{1}{2}t^2(f''(\theta) + f''(\theta')) > f(u)+f(v)$$
where the last inequality follows from the positivity of $f''$.
\hfill$\Box$

{\em Proof of Lemma~\ref{ultralem}.}
Since the branch lengths of $t$ are clock-like, it follows that $\kk$ and hence $\tau$ is an ultrametric, i.e. for any three leaves of $t$ we have: $$\tau_{ij} \leq \max\{\tau_{ik}, \tau_{jk}\}.$$  It follows that $\tau^{\rho}$ (where $\rho = k/k'$) satisfies precisely the same  ultrametric conditions as $\tau$ and so we can assign (unique) positive branch lengths to $t$ that realize
$\tau^\rho$ and which are clock-like. From (\ref{thishelps}), these branch lengths satisfy (\ref{ultra1}) of Lemma~\ref{ultralem}.  Moreover, since $t$ has at least two leaves, we can select two leaves (say $u, v$) so that the path connecting $u$ and $v$ contains $\rho$.  Then $l_{ij} = 2h(t, l)$, and $l'_{ij} = 2h(t, l')$.  Now
$\tau'_{ij} = (\tau_{ij})^\rho$ and so, by  (\ref{taudef}) and (\ref{taudef2}), we have:
$$(1+ \frac{2h(t, l)}{k\gamma})^\rho = 1+\frac{2h(t,l')}{k'\gamma},$$
from which equality (\ref{ultra2}) of Lemma~\ref{ultralem} now follows.
\hfill$\Box$

\end{document}